\RequirePackage{filecontents}

\RequirePackage{snapshot}
\documentclass[aps,onecolumn,notitlepage,secnumarabic,superscriptaddress]{preprint}
\def\Preprint{}
\renewcommand\documentclass[2][]{}

\documentclass[a4paper,intlimits]{spie}

\usepackage[T1]{fontenc}
\usepackage{amsmath}
\usepackage{bm}
\usepackage{txfonts}
\makeatletter
\renewcommand*{\vec}[1]{\bm{#1}}
\makeatother
\usepackage{graphicx}
\graphicspath{{figures/}}
\usepackage[american]{babel}
\usepackage{microtype}
\usepackage{textcomp}
\usepackage{braket}
\usepackage{booktabs}
\usepackage{rotating}
\usepackage[pdfborder={0 0 0.5}]{hyperref}

\newcommand*{\I}{\mathrm{i}}

\newcommand*{\e}{\mathrm{e}}
\newcommand*{\commut}[2]{%
  \mathchoice{\left[#1,#2\right]}{[#1,#2]}{[#1,#2]}{[#1,#2]}%
}
\newcommand*{\acommut}[2]{%
  \mathchoice{\left\{#1,#2\right\}}{\{#1,#2\}}{\{#1,#2\}}{\{#1,#2\}}%
}

\newcommand*{\op}[1]{\hat{#1}}
\newcommand*{\opHo}{\op{H}_0}
\newcommand*{\opHC}{\op{H}_C}
\newcommand*{\opp}{\op{\vec p}}
\newcommand*{\oppo}{\op{p}_0}
\newcommand*{\opsigma}[1][]{\ifthenelse{\equal{#1}{}}{\vec{\op{\sigma}}}{\op{\sigma}_{#1}}}
\newcommand*{\opSigma}[1][]{\ifthenelse{\equal{#1}{}}{\vec{\op{\Sigma}}}{\op{\Sigma}_{#1}}}
\newcommand*{\opS}[1][]{\ifthenelse{\equal{#1}{}}{\op{\vec{S}}}{\op{S}_{#1}}}
\newcommand*{\opL}[1][]{\ifthenelse{\equal{#1}{}}{\op{\vec{L}}}{\op{L}_{#1}}}
\newcommand*{\opJ}[1][]{\ifthenelse{\equal{#1}{}}{\op{\vec{J}}}{\op{J}_{#1}}}
\newcommand*{\opN}[1][]{\ifthenelse{\equal{#1}{}}{\op{\vec{N}}}{\op{N}_{#1}}}
\newcommand*{\opW}[1][]{\ifthenelse{\equal{#1}{}}{\op{\vec{W}}}{\op{W}_{#1}}}
\newcommand*{\opSP}[1][]{\ifthenelse{\equal{#1}{}}{\vec{\op{S}}_\mathrm{P}}{\op{S}_{\mathrm{P},#1}}}
\newcommand*{\opSPnr}[1][]{\ifthenelse{\equal{#1}{}}{\vec{\op{S}}_\mathrm{P,nr}}{\op{S}_{\mathrm{P,nr},#1}}}
\newcommand*{\opSFW}[1][]{\ifthenelse{\equal{#1}{}}{\vec{\op{S}}_\mathrm{FW}}{\op{S}_{\mathrm{FW},#1}}}
\newcommand*{\opSCz}[1][]{\ifthenelse{\equal{#1}{}}{\vec{\op{S}}_\mathrm{Cz}}{\op{S}_{\mathrm{Cz},#1}}}
\newcommand*{\opSF}[1][]{\ifthenelse{\equal{#1}{}}{\vec{\op{S}}_\mathrm{F}}{\op{S}_{\mathrm{F},#1}}}
\newcommand*{\opSCh}[1][]{\ifthenelse{\equal{#1}{}}{\vec{\op{S}}_\mathrm{Ch}}{\op{S}_{\mathrm{Ch},#1}}}
\newcommand*{\opSPr}[1][]{\ifthenelse{\equal{#1}{}}{\vec{\op{S}}_\mathrm{Pr}}{\op{S}_{\mathrm{Pr},#1}}}
\newcommand*{\opSFG}[1][]{\ifthenelse{\equal{#1}{}}{\op{\vec{S}}_\mathrm{FG}}{\op{S}_{\mathrm{FG},#1}}}

\newcommand*{\trans}[1]{#1^{\mathsf{T}}}
\newcommand*{\htrans}[1]{#1^{\dagger}}
\newcommand*{\GAMMA}{\mathrm{\Gamma}}

\begin{document} 

\title{Spin dynamics in relativistic light-matter interaction}  

\ifdefined\Preprint 
\author{Heiko Bauke}
\email{heiko.bauke@mpi-hd.mpg.de}
\affiliation{Max-Planck-Institut f{\"u}r Kernphysik, Saupfercheckweg~1,
  69117~Heidelberg, Germany} 
\author{Sven Ahrens}
\thanks{Current affiliation: Beijing Computational Science Research Center, Beijing 100094, China}
\affiliation{Max-Planck-Institut f{\"u}r Kernphysik, Saupfercheckweg~1,
  69117~Heidelberg, Germany} 
\affiliation{{Intense Laser Physics Theory Unit and Department of
    Physics, Illinois State University, Normal, Illinois 61790-4560 USA}} 
\author{Christoph~H.~Keitel}
\affiliation{Max-Planck-Institut f{\"u}r Kernphysik, Saupfercheckweg~1,
  69117~Heidelberg, Germany}
\author{Rainer Grobe}
\affiliation{Max-Planck-Institut f{\"u}r Kernphysik, Saupfercheckweg~1,
  69117~Heidelberg, Germany} 
\affiliation{{Intense Laser Physics Theory Unit and Department of
    Physics, Illinois State University, Normal, Illinois 61790-4560 USA}} 
\else
\author{Heiko~Bauke\supit{a}, 
  Sven~Ahrens\supit{a,b}\thanks{\small\hspace{1ex}Current affiliation: Beijing Computational Science
  Research Center, Beijing 100094, China}, 
  Christoph~H.~Keitel\supit{a},
  Rainer~Grobe\supit{a,b}
  \skiplinehalf
  \supit{a}Max-Planck-Institut f{\"u}r Kernphysik,
  Saupfercheckweg~1,~69117~Heidelberg, Germany;\\[0.4ex]
  \supit{b}Intense Laser Physics Theory Unit and Department of
  Physics, Illinois State University, Normal, Illinois 61790-4560 USA
}

\authorinfo{Corresponding author: Heiko Bauke,
  e-mail: heiko.bauke@mpi-hd.mpg.de}

\maketitle 

\fi

\begin{abstract}
  Various spin effects are expected to become observable in light-matter
  interaction at relativistic intensities.  Relativistic quantum
  mechanics equipped with a suitable relativistic spin operator forms
  the theoretical foundation for describing these effects.  Various
  proposals for relativistic spin operators have been offered by
  different authors, which are presented in a unified way.  As a result
  of the operators' mathematical properties only the Foldy-Wouthuysen
  operator and the Pryce operator qualify as possible proper
  relativistic spin operators.  The ground states of highly charged
  hydrogen-like ions can be utilized to identify a legitimate
  relativistic spin operator experimentally.  Subsequently, the
  Foldy-Wouthuysen spin operator is employed to study electron-spin
  precession in high-intensity standing light waves with elliptical
  polarization.  For a correct theoretical description of the predicted
  electron-spin precession relativistic effects due to the spin angular
  momentum of the electromagnetic wave has to be taken into account even
  in the limit of low intensities.
\end{abstract}

\keywords{relativistic quantum dynamics, spin, light-matter interaction}

\ifdefined\Preprint 
\maketitle 
\else
\fi

\section{Introduction}
\label{sec:intro} 

Employing novel light sources such as the ELI-Ultra High Field Facility,
for example, that envisage to provide field intensities in excess of
$10^{20}\,\mathrm{W/cm^2}$ and field frequencies in the x-ray domain
\cite{Altarell:etal:2007:XFEL,Yanovsky:Chvykov:etal:2008:Ultra-high_intensity-,McNeil:Thompson:2010:X-ray_free-electron,Emma:Akre:etal:2010:First_lasing,Mourou:Fisch:etal:2012:Exawatt-Zettawatt_pulse}
light-matter interaction in the relativistic regime may be probed
experimentally.  Relativistic quantum mechanics predicts various new
phenomena to occur in this regime
\cite{Ehlotzky:Krajewska:Kaminski:2009:Fundamental_processes,Di-Piazza:Muller:etal:2012:Extremely_high-intensity},
for example, multiphoton scattering, radiation reaction effects,
vacuum-polarization effects or even pair creation
\cite{Blaschke:Prozorkevich:etal:2006:Pair_Production,Bell:Kirk:2008:Possibility_of,Pike:Mackenroth:etal:2014:photon-photon_collider}.
Furthermore, electrons in strong electromagnetic fields can exhibit
distinct spin effects
\cite{Walser:Urbach:etal:2002:Spin_and,Faisal:Bhattacharyya:2004:Spin_Asymmetry,Brodin:Marklund:etal:2011:Spin_and,Ahrens:Bauke:etal:2012:Spin_Dynamics,Ahrens:Muller:etal:2013:Spin_effects,Klaiber:Yakaboylu:etal:2014:Spin_dynamics}.

Relativistic quantum mechanics has to be employed to study spin
phenomena in strong electromagnetic fields.  According to the formalism
of quantum mechanics, each measurable quantity---as the spin, for
example---is represented by a Hermitian operator.  However, there is no
universally accepted operator to describe an electron's spin degree of
freedom within the framework of relativistic quantum mechanics.  Thus,
we first investigate the properties of different proposals for a
relativistic spin operator and show that most candidates are lacking
essential features of proper angular momentum operators
\cite{Bauke:etal:2014:What_is,Bauke:Ahrens:etal:2014:Relativistic_spin}.
Only the so-called Foldy-Wouthuysen and the Pryce operators qualify as
proper relativistic spin operators.  The various spin operators predict
different expectation values when electrons interact with
electromagnetic potentials.  In this way, one may distinguish between
the proposed relativistic spin operators by experimental means.  In
particular, eigenstates of highly charged hydrogen-like ions may be
utilized to identify a legitimate relativistic spin operator
experimentally.

A further relativistic spin phenomenon, which we study in more detail,
is the coupling of the spin angular momentum of light beams with
elliptical polarization to the spin degree of freedom of free electrons
\cite{Bauke:Ahrens:etal:2014:Electron-spin_dynamics,Bauke:Ahrens:etal:2014:Electron-spin_dynamics_2}.
This coupling, which is of similar origin as the well-known spin-orbit
coupling, and the magnetic field lead to electron-spin precession.  The
spin-precession frequency is proportional to the product of the
laser-field's intensity and its spin density.  To derive the correct
spin-precession frequency relativistic corrections to the
nonrelativistic Pauli equation, which account for the light's spin
density, have to be taken into account.  The quantum mechanical
interactions of the electron's spin with the laser's rotating magnetic
field, which may be characterized by the nonrelativistic Pauli equation,
and the electron spin's interaction with the laser field's spin density,
which results via the relativistic corrections, counteract each other.
As a result, a net electron-spin rotation remains with a precession
frequency that is much smaller than the frequency predicted by a
nonrelativistic theory.  These relativistic effects are maintained even
if the involved electromagnetic field strengths are nonrelativistic.

\section{Relativistic spin operators}
\label{sec:relativistic_spin}

A Lorentz invariant quantum mechanical description of the motion of an
electron in electromagnetic fields is provided by the time-dependent
Dirac equation.  For a particle of rest mass $m_0$ and charge $q$ it is
given by (units are used in this section for which $\hbar=1$)
\begin{equation}
  \label{eq:Dirac_equation}
  \I\frac{\partial \Psi(\vec r, t)}{\partial t} =
  \op H\Psi(\vec r, t) = 
  \left(
    c\vec\alpha\cdot\left(\opp- q \vec A(\vec r, t) \right) + 
    q \phi(\vec r, t) + m_0c^2\beta
  \right)\Psi(\vec r, t)\,,
\end{equation}
with the electromagnetic potentials $\phi(\vec r, t)$ and $\vec A(\vec
r, t)$, the speed of light $c$, the canonical momentum operator
$\opp=-\I\vec\nabla$, and the matrices $\vec\alpha=\trans{(\alpha_1,
  \alpha_2, \alpha_3)}$ and $\beta$. These $4\times4$ matrices obey the
algebra
\begin{equation}
  \label{eq:Dirac_algebra}
  \alpha_i^2 = \beta^2 = 1\,,\qquad
  \alpha_i\alpha_k + \alpha_k\alpha_i = 2\delta_{i,k}\,,\qquad
  \alpha_i\beta + \beta\alpha_i = 0\,.
\end{equation}
To specify our notation and abbreviations, we employ the Dirac
representation, where the matrices $\alpha_i$ and $\beta$ are defined as
\begin{equation}
  \alpha_i =
  \begin{pmatrix}
    0 & \sigma_i \\ \sigma_i & 0
  \end{pmatrix}\quad\text{for $i=1,2,3$}\,,
  \qquad
  \beta =
  \begin{pmatrix}
    \mathbb{I}_2 & 0 \\ 0 & -\mathbb{I}_2
  \end{pmatrix}
\end{equation}
in terms of the three $2\times2$ Pauli matrices
$\vec\sigma=\trans{(\sigma_1, \sigma_2, \sigma_3)}$.  In the Dirac
representation, the Pauli matrices are given by
\begin{equation}
  \sigma_1 = \begin{pmatrix} 0 & 1 \\ 1 & 0 \end{pmatrix} \,,\qquad
  \sigma_2 = \begin{pmatrix} 0 & -\I \\ \I & 0 \end{pmatrix} \,,\qquad
  \sigma_3 = \begin{pmatrix} 1 & 0 \\ 0 & -1 \end{pmatrix} \,.
\end{equation}
The symbol $\mathbb{I}_2$ denotes the $2\times2$ identity matrix.  The
free-particle Dirac Hamiltonian with $\vec A(\vec r, t)=0$ and
$\phi(\vec r, t)=0$ will be denoted by $\opHo$.  The free-particle Dirac
Hamiltonian features positive as well as negative energy-eigenvalues.
We will also use the operator $\op p_0$ to denote
\begin{equation}
  \op p_0 = \sqrt{\opHo^2/c^2} = \sqrt{ m_0^2c^2 + \opp^2} \,,
\end{equation}
which has the same eigenstates as $\opHo$ but all its eigenvalues being
positive. Furthermore, the matrices 
\begin{equation}
  \Sigma_i =
  \begin{pmatrix}
    \sigma_i & 0 \\ 0 & \sigma_i 
  \end{pmatrix}\quad\text{for $i=1,2,3$}
\end{equation}
will be employed, which will be commonly combined into the
three-component operator $\vec\Sigma=\trans{(\Sigma_1, \Sigma_2,
  \Sigma_3)}$.

A relativistic spin operator may be introduced by splitting the
undisputed total angular momentum operator $\opJ$, which is in the Dirac
representation
\begin{equation}
  \opJ = \vec r \times (-\I\vec\nabla) + \frac{1}{2}\opSigma \,,
\end{equation}
into an external part $\opL$ and an internal part $\opS$, viz.,
$\opJ=\opL+\opS$.  These parts are commonly referred to as the orbital
angular momentum and the spin.  Because the orbital angular momentum
operator $\opL$ is related to the position operator $\vec{\op{r}}$ and
the momentum operator $\opp=-\I\vec\nabla$ via
$\opL=\vec{\op{r}}\times\opp$, different definitions of the spin
operator $\vec{\op S}$ imply different relativistic position operators
$\vec{\op r}$.  Thus, the question for the right splitting of the total
angular momentum into an orbital part and a spin part is closely related
to the quest for the right relativistic position operator
\cite{Newton:Wigner:1949:Localized_States,Jordan:Mukunda:1963:Lorentz-Covariant_Position_Operators,OConnell:Wigner:1978:Position_operators}.

\begin{sidewaystable}
  \centering
  \caption{Brief summary of in the literature proposed spin operators'
    definitions and their mathematical properties adopted from {\em
      Phys. Rev. A}~{\bf 89}(5), 052101 (2014). The table indicates from
    left to right the name and the definition of the various spin
    operators, if they are Hermitian, if they transform under rotations
    like a vector, if they commute with the free Dirac Hamiltonian, if
    they obey the angular momentum algebra, if eigenvalues are $\pm1/2$,
    and if the operators are related to the Pauli spin operator $\opSP$
    via an orthogonal transformation.  The corresponding orthogonal
    transformations are given in \eqref{eq:UFW}, \eqref{eq:UPr} and in
    {\em Phys. Rev. A}~{\bf 89}(5), 052101 (2014).}
  \label{tab:spin_operators}
  \setlength{\tabcolsep}{0.75ex}%
  \setlength{\cmidrulekern}{0.5ex}%
  \vspace*{0.5ex}%
  \renewcommand{\arraystretch}{1.67}
  \begin{tabular}{@{}llcccccc@{}}
    \toprule
    operator name & 
    definition & $\opS=\htrans{\opS}$ & 
    \parbox{11ex}{\centering $\commut{\op{J}_i}{\opS[j]}=$ $\I\varepsilon_{i,j,k}\opS[k]$} &
    $\commut{\opHo}{\opS}=0$ &
    \parbox{11ex}{\centering $\commut{\opS[i]}{\opS[j]}=$ $\I\varepsilon_{i,j,k}\opS[k]$} &
    \parbox{11ex}{\centering eigenvalues equal $\pm1/2$} & $\opS = \op{T}\opSP\op{T}^{-1}$ \\
    \cmidrule(r){1-1}\cmidrule(lr){2-2}\cmidrule(lr){3-3}\cmidrule(lr){4-4}\cmidrule(lr){5-5}\cmidrule(lr){6-6}\cmidrule(lr){7-7}\cmidrule(l){8-8}
      Pauli op.
      \cite{Hill:Landshoff:1938,Dirac:1958:principles,Dirac:1971:Positive-Energy,Ohanian:1986:spin,Lifshitz:1996:QED,Feynman:1998:QED} & 
      $\opSP = \dfrac 12 \opSigma$ &
      yes & yes & no & yes & yes & --- \\[2.50ex]
      Foldy-Wouthuysen op.
      \cite{Foldy:Wouthuysen:1950:Non-Relativistic_Limit,Vries:1970:Foldy-Wouthuysen_Transformations,Costella:McKellar:1995:Foldy-Wouthuysen,Schweber:2005:qft,Caban:etal:2013:A_spin_observable} & 
      $ \opSFW = \dfrac{1}{2}\opSigma + 
      \dfrac{\I\beta}{2\oppo}\op{\vec{p}}\times\vec\alpha -
      \dfrac{\op{\vec{p}}\times(\opSigma\times\op{\vec{p}})}{2\oppo(\oppo+m_0c)}$ & 
      yes & yes & yes & yes & yes & yes \\[2.50ex]
      Czachor op. \cite{Czachor:1997:relativistic_EPR} & 
      $\opSCz 
      = \dfrac{m_0^2c^2}{2\oppo^2}\opSigma + 
      \dfrac{\I m_0c\beta}{2\oppo^2}\op{\vec{p}}\times\vec\alpha +
      \dfrac{\op{\vec{p}}\cdot\opSigma}{2\oppo^2}\op{\vec{p}}$ & 
      yes & yes & yes & no & no & no \\[2.50ex]
      Frenkel op.
      \cite{Hilgevoord:Wouthuysen:1963:Dirac_spin,Wightman:1960,Bargmann:etal:1959:Precession} & 
      $\opSF = \dfrac{1}{2}\opSigma +
      \dfrac{\I\beta}{2m_0c}\op{\vec{p}}\times\vec\alpha$ & 
      yes & yes & yes & no & no & no \\[2.50ex]
      Chakrabarti op.
      \cite{Chakrabarti:1963:Canonical_Form,Gursey:1965,Gursey:1965:Equivalent_formulations,Choi:2013:Relativistic_spin_operator} & 
      $ \opSCh = 
      \dfrac{1}{2}\opSigma + 
      \dfrac{\I}{2m_0c}\vec\alpha\times\op{\vec{p}} +
      \dfrac{\op{\vec{p}}\times(\opSigma\times\op{\vec{p}})}{2m_0c(m_0c+\oppo)}$ &
      no & yes & no & yes & yes & yes \\[2.50ex]
      Pryce op.
      \cite{Pryce:1935:Commuting_Co-Ordinates,Pryce:1948:Mass-Centre,Macfarlane:1963:Relativistic_Two_Particle,Berg:1965:Position_Spin,Ryder:1999:Relativistic_Spin} & 
      $ \opSPr = 
      \dfrac{1}{2}\beta\opSigma + 
      \dfrac{1}{2}\opSigma\cdot\opp(1-\beta)\dfrac{\opp}{\opp^2}$ & 
      yes & yes & yes & yes & yes & yes \\[2.50ex]
      Fradkin-Good op. 
      \cite{Fradkin:Good:1961:Polarization_Operators,Kirsch:Ryder:Hehl:2001:Gordon_decompositions} & 
      $ \opSFG = 
      \dfrac{1}{2}\beta\opSigma + 
      \dfrac{1}{2}\opSigma\cdot\opp\left(\dfrac{\opHo}{c\oppo}-\beta\right)\dfrac{\opp}{\opp^2}$ & 
      yes & yes & yes & no & yes & no \\
    \bottomrule
    \end{tabular}
\end{sidewaystable}

There are two complementary approaches to determine a suitable
relativistic spin operator from a set of possible candidates.  A common
approach is to judge a spin operator candidate by its mathematical
properties.  This means to analyze its symmetries, its behavior under
various transformations, its algebraic features, and so on.
Historically, most relativistic spin operators have been proposed on
purely mathematical arguments.  Considering that relativistic spin
effects are expected to be detectable at high-intensity laser facilities
one may also compare theoretical predictions based on various
relativistic spin operators to experimental measurements.  In this way,
candidates for relativistic spin operators that are not compatible with
measurements can be ruled out on the basis of experimental results,
rather than solely by mathematical reasoning.  Here, we will adopt both
approaches.

Mathematically, we demand from a proper relativistic spin operator
$\opS=\trans{(\opS[1],\opS[2],\opS[3])}$ and its three components the
following features:
\begin{enumerate}
\item Each component of a spin operator should be a Hermitian operator.
\item The physical quantity that is represented by the operator $\opS$
  should not depend on the orientation of the chosen coordinate
  system. Thus, a spin operator must transform under rotations like a
  vector, which is ensured by fulfilling
  \cite{Sakurai:Napolitano:2010:Modern_QM}
  \begin{equation}
    \label{eq:commut_J_S}
    \textstyle\commut{\op{J}_i}{\opS[j]} =
    \I\varepsilon_{i,j,k}\opS[k]
  \end{equation}
  with $\varepsilon_{i,j,k}$ denoting the Levi-Civita symbol.
\item It is also required to commute with the free Dirac Hamiltonian,
  i.\,e., $\commut{\opHo}{\opS}=0$.  This property ensures that the
  relativistic spin operator is a constant of motion if forces are
  absent, such that spurious Zitterbewegung of the spin is prevented.
\item A spin operator must feature the two eigenvalues $\pm1/2$ and it
  has to obey the angular momentum algebra
  \begin{equation}
    \label{eq:commut_S}
    \textstyle\commut{\opS[i]}{\opS[j]}=\I\varepsilon_{i,j,k}\opS[k] \,.
  \end{equation}
  These two requirements are commonly regarded as \emph{the} fundamental
  properties of angular momentum operators of spin one-half particles
  \cite{Sakurai:Napolitano:2010:Modern_QM}.
\end{enumerate}

\noindent%
Table~\ref{tab:spin_operators} gives an overview over several spin
operators, which have been proposed in the literature, and presents
their mathematical key features.  A detailed description of these
operators and their relations to each other is given elsewhere
\cite{Bauke:Ahrens:etal:2014:Relativistic_spin}.  Note that some of
these spin operators have been discovered by different authors in
different contexts yielding different but mathematically equivalent
forms.  For example, the so-called Newton-Wigner spin operator
\cite{Newton:Wigner:1949:Localized_States}, which may be written as
\cite{Caban:Rembielinski:etal:2013:Spin_operator}
\begin{equation}
  \label{eq:SFW_bar_bar_bar}
  \opSFW = \frac{\oppo}{2m_0c}\opSigma - 
  \frac{\opp\cdot\opSigma}{2m_0c(m_0c+\oppo)}\opp - 
  \opp\times\vec\alpha\frac{\I\opHo}{2m_0c^2\oppo}\,,
\end{equation}
is just another way to express the Foldy-Wouthuysen spin operator given
in Tab.~\ref{tab:spin_operators}.

The Foldy-Wouthuysen, the Chakrabarti, the Pryce, and the Fradkin-Good
spin operators are equivalent in the positive-energy subspace of
free-particle states.  This can be shown easily in the Foldy-Wouthuysen
representation.  The transition from the standard representation, where
the Dirac Hamiltonian has the form \eqref{eq:Dirac_equation} and the
various spin operators have the forms given in
Tab.~\ref{tab:spin_operators}, is mediated via the nonlocal unitary
transform
\begin{equation}
  \label{eq:UFW}
  \op{T}_\mathrm{FW} = 
  \frac{\oppo + m_0c -\beta\vec{\alpha}\cdot\opp}
  {\sqrt{2\oppo(\oppo + m_0c)}}\,.
\end{equation}
In the Foldy-Wouthuysen representation the free-particle Dirac
Hamiltonian becomes diagonal,
\begin{equation}
  \opHo' = \op{T}_\mathrm{FW}^{-1}\opHo\op{T}_\mathrm{FW} = c\beta\oppo
\end{equation}
and the Foldy-Wouthuysen spin operator has the form
\begin{equation}
  \label{eq:SFW_FW}
  \opSFW' =
  \op{T}_\mathrm{FW}^{-1}\opSFW\op{T}_\mathrm{FW} = 
  \begin{pmatrix}
     \frac 1 2\vec\sigma & 0 \\[1ex]
    0 &  \frac 1 2\vec\sigma
  \end{pmatrix}\,.
\end{equation}
Since the momentum operator is invariant under the Foldy-Wouthuysen
transformation $\op{T}_\mathrm{FW}$, the simultaneous eigenstates of the
free-particle Dirac Hamiltonian, the momentum operator, and the $z$
component of the Foldy-Wouthuysen spin operator are in the
Foldy-Wouthuysen representation given by 
\begin{align}
  \label{eq:free_FW}
  \vec s_{\mathrm{FW},+,\vec p, \uparrow} & = 
  \begin{pmatrix}1\\0\\0\\0\end{pmatrix}\e^{\I\vec p\cdot\,\vec r}\,, &
  \vec s_{\mathrm{FW},+,\vec p, \downarrow} & = 
  \begin{pmatrix}0\\1\\0\\0\end{pmatrix}\e^{\I\vec p\cdot\,\vec r}\,, &
  \vec s_{\mathrm{FW},-,\vec p, \uparrow} & = 
  \begin{pmatrix}0\\0\\1\\0\end{pmatrix}\e^{\I\vec p\cdot\,\vec r}\,, &
  \vec s_{\mathrm{FW},-,\vec p, \downarrow} & = 
  \begin{pmatrix}0\\0\\0\\1\end{pmatrix}\e^{\I\vec p\cdot\,\vec r}\,,
\end{align}
where the indices indicate the sign of the energy eigenvalue, the
momentum eigenvalues and the spin eigenvalue ($\uparrow$~for~$1/2$,
$\downarrow$~for~$-1/2$).  In the Foldy-Wouthuysen representation the
Chakrabarti, the Pryce, and the Fradkin-Good spin operators are given by
the rather simple expressions
\begin{align}
  \label{eq:SCh_FW}
  \opSCh' & =
  \op{T}_\mathrm{FW}^{-1}\opSCh\op{T}_\mathrm{FW} = 
  \begin{pmatrix}
    \frac 1 2\vec\sigma & \I\frac{\vec\sigma\times\vec\opp}{m_0c} \\[1ex]
    0 & \frac 1 2\vec\sigma
  \end{pmatrix}\,, \\
  \label{eq:SPr_FW}
  \opSPr' & =
  \op{T}_\mathrm{FW}^{-1}\opSPr\op{T}_\mathrm{FW} = 
  \begin{pmatrix}
     \frac 1 2\vec\sigma & 0 \\[1ex]
     0 & -\frac 1 2\vec\sigma + \frac{\vec\sigma\cdot\,\vec\opp}{\opp^2}\opp
  \end{pmatrix}\,,\\
  \label{eq:SFG_FW}
  \opSFG' & =
  \op{T}_\mathrm{FW}^{-1}\opSFG\op{T}_\mathrm{FW} = 
  \begin{pmatrix}
     \frac 1 2\vec\sigma & 0 \\[1ex]
    0 & -\frac 1 2\vec\sigma
  \end{pmatrix}\,.
\end{align}
Because the operators \eqref{eq:SFW_FW}, \eqref{eq:SCh_FW},
\eqref{eq:SPr_FW}, and \eqref{eq:SFG_FW} have the same upper left
$2\times2$ matrix, all these operators act in the same way on the
positive-energy free particle states given in \eqref{eq:free_FW}.  For
quantum states that are superpositions of free-particle states with
positive and negative energy, however, these spin operators are not
equivalent.

On the basis of the four criteria given above, one may argue that only
the Foldy-Wouthuysen spin operator and the Pryce spin operator qualify
as proper relativistic spin operators because only these two fulfill all
four criteria.  However, the question of which of the proposed
relativistic spin operators in Tab.~\ref{tab:spin_operators} provides
the correct mathematical description of spin can be answered definitely
only by comparing theoretical predictions with experimental results.
Because if an interaction with some external fields is introduced a
superposition of positive-energy free-particle states evolves such that
negative-energy free-particle states become populated.  The proposed
spin operators are not equivalent if they are applied to positive-energy
states of Hamiltonians with nonvanishing electromagnetic fields.
Therefore, it becomes possible to distinguish between the various spin
operators by determining their expectation values for electrons
interacting with electromagnetic fields.

For this purpose it is desirable to employ a physical system that shows
strong relativistic effects and is as simple as possible.  Such a setup
is provided by the bound eigenstates of highly charged hydrogen-like
ions, i.\,e., atomic systems with an atomic core of $Z$ protons and a
single electronic charge.  These ions can be produced at storage rings
\cite{Stohlker:etal:1993:Ground-state_Lamb_shift} or by utilizing
electron beam ion traps
\cite{Robbins:etal:2006:Polarization_measurements,Kluge:etal:2008:HITRAP}
up to $Z=92$ (hydrogen-like uranium).  The degenerate bound eigenstates
of the corresponding Coulomb-Dirac Hamiltonian (in atomic units)
\begin{equation}
  \opHC = \opHo - \frac{Z}{|\vec r|} 
\end{equation}
are commonly expressed as simultaneous eigenstates
$\psi_{n,j,m,\kappa}$ of $\opHC$, $\opJ^2$, $\opJ[3]$, and the
so-called spin-orbit operator $\op{K}=\mbox{$\beta\{\opSigma\cdot[\vec
  r\times(-\I\vec\nabla)+1)]\}$}$ fulfilling the eigenequations
\cite{Bethe:Salpeter:2008:Atoms,Thaller:2005:vis_QM_II}
\begin{subequations}
  \begin{align}
    \opHC \psi_{n,\kappa,j,m} & = \mathcal{E}(n, \kappa)\psi_{n,\kappa,j,m}\,, &
    n & = 1,2,\dots\,,\\
    \op{K} \psi_{n,\kappa,j,m} & = \kappa\psi_{n,\kappa,j,m}\,, & 
    |\kappa| & = 1,2,\dots,n \,,\kappa\ne-n\,,\\
    \opJ^2 \psi_{n,\kappa,j,m} & = j(j+1)\psi_{n,\kappa,j,m}\,, &
    j & = |\kappa|-\tfrac{1}{2} \,,\\
    \opJ[3] \psi_{n,\kappa,j,m} & = m\psi_{n,\kappa,j,m} \,, &
    m & = -j,(j-1),\dots,j \,.
  \end{align}
\end{subequations}
The eigenenergies are given with $\alpha_\mathrm{el}$ denoting the fine
structure constant by
\begin{equation}
  \label{eq:energy}
  \mathcal{E}(n, j) = 
  m_0c^2\left[
    1 + \left(
      \dfrac{\alpha_\mathrm{el}^2Z^2}
      {n-j-{1}/{2}+\sqrt{\vphantom{1_1}\smash[b]{(j-{1}/{2})^2-\alpha^2_\mathrm{el}Z^2}}}
    \right)
  \right]^{-{1}/{2}}\,.
\end{equation}
The degenerate hydrogenic ground state is with
$\gamma=\sqrt{\vphantom{1_1}\smash[b]{1-Z^2\alpha_\mathrm{el}^2}}$, the
radial function
\begin{equation}
  \psi(r) = 
  \frac{\e^{-m_0Zr} }{(2m_0Zr)^{1-\gamma}}\,,
\end{equation}
and the normalizing factor 
\begin{equation}
  \mathcal{N}=
  (2m_0Z)^{3/2}
  \sqrt{\frac{1+\gamma}{2\GAMMA(1+2\gamma)}} 
\end{equation}
given by the two wave functions \cite{Bjorken:Drell:1964:RQM}
\begin{subequations}
  \label{eq:hydrogen_real}
  \begin{align}
    \label{eq:hydrogen_real_1}
    \psi_{1, 1, \frac{1}{2}, \frac{1}{2}}(r,\theta,\phi) & = \mathcal{N}\psi(r)
    \begin{pmatrix}
      Y_{0,0}(\theta,\phi) \\
      0 \\
      \I\frac{1-\gamma}{Z\alpha_\mathrm{el}}\sqrt{\frac13}Y_{1,0}(\theta,\phi) \\[1.5ex]
      -\I\frac{1-\gamma}{Z\alpha_\mathrm{el}}\sqrt{\frac23}Y_{1,1}(\theta,\phi)
    \end{pmatrix}\,,\\
  \label{eq:hydrogen_real_2}
    \psi_{1, 1, \frac{1}{2}, -\frac{1}{2}}(r,\theta,\phi) & = \mathcal{N}\psi(r)
    \begin{pmatrix}
      0 \\
      Y_{0,0}(\theta,\phi) \\
      \I\frac{1-\gamma}{Z\alpha_\mathrm{el}}\sqrt{\frac23}Y_{1,-1}(\theta,\phi) \\[1.5ex]
      -\I\frac{1-\gamma}{Z\alpha_\mathrm{el}}\sqrt{\frac13}Y_{1,0}(\theta,\phi)
    \end{pmatrix}\,.
  \end{align}
\end{subequations}

\begin{figure}
  \centering
  \includegraphics[scale=0.5]{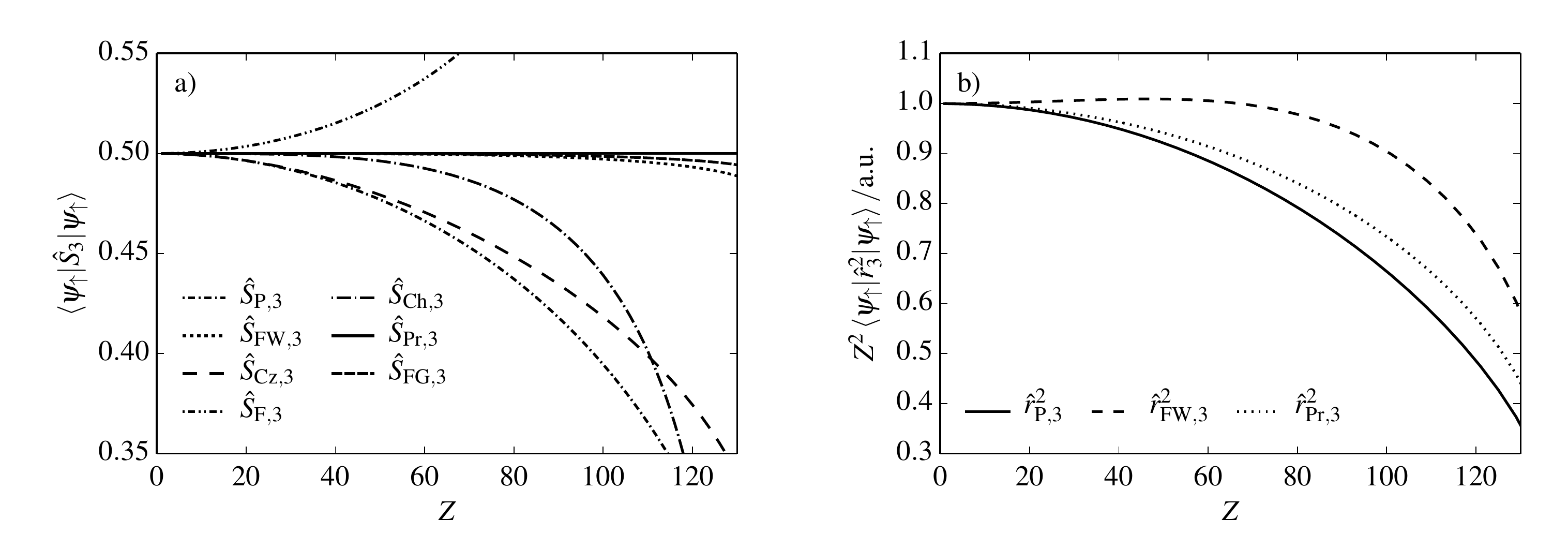}
  \caption{(a) Spin expectation values of various relativistic spin
    operators for the hydrogenic ground state (\ref{eq:hydrogen_real_1})
    as a function of the atomic number $Z$, adopted from {\em New
      J. Phys.}~{\bf 16}(4), 043012 (2014).  For the ground state
    (\ref{eq:hydrogen_real_2}) we find the same spin expectation values
    but with opposite sign (not displayed in the plot). (b) Expectation
    values of the position variance with respect to the $z$ coordinate
    for the hydrogenic ground state (\ref{eq:hydrogen_real_1}) as a
    function of the atomic number $Z$.  The wave function's variance
    depends on the definition of the relativistic position operator,
    which is induced by the definition of the relativistic spin
    operator.}
  \label{fig:spin}
\end{figure}

The spin expectation values of the $z$ component of the in
Tab.~\ref{tab:spin_operators} defined spin operators are presented in
Fig.~\ref{fig:spin}\,a).  For small atomic numbers ($Z\lessapprox20$),
all spin operators yield about $1/2$; for larger $Z$ when relativistic
effects set in, however, expectation values differ significantly from
each other.  While for Pauli, Fouldy-Wouthuysen, Czachor, Chakrabarti,
and Fradkin-Good spin operators the spin expectation value is reduced,
the expectation value of the Frenkel spin operator exceeds $1/2$.  Only
for the Pryce operator we find that the spin expectation values is $1/2$
for all values of $Z$, which can be also shown analytically
\cite{Bauke:Ahrens:etal:2014:Relativistic_spin}.  A comparison of the
results of a spin-measurement experiment for groundstate electrons in
hydrogen-like highly-charged ions to the numerical results
Fig.~\ref{fig:spin}\,a) would allow to find a suitable relativistic spin
operator or at least to rule out some candidates
\cite{Bauke:etal:2014:What_is}.

As outlined above, each definition of a relativistic spin operator
$\opS$ induces also a relativistic position operator
$\vec{\op{r}}$. Because only the Fouldy-Wouthuysen and the Pryce
operators fulfill all our mathematical criteria for a proper spin
operator we focus on these two in the following.  For determining the
corresponding position operator the defining relation
\begin{equation}
  \opJ=\vec{r}\times(-\I\vec\nabla) + \vec\Sigma/2 =
  \op{\vec{r}}\times(-\I\vec\nabla)+\opS
\end{equation}
is rather unwieldy.  In the case of the Fouldy-Wouthuysen and the Pryce
operators it is more convenient to utilize the fact that these are
related to the Pauli spin operator via a unitary transform.  The
position operator of the Pauli spin operator is just
\begin{equation}
  \op{\vec r}_\mathrm{P} = \vec r\,.
\end{equation}
Therefore, the position operators of the Fouldy-Wouthuysen and the Pryce
operators are given by
\begin{equation}
  \label{eq:r_FW}
  \op{\vec r}_\mathrm{FW} = 
  \op{T}_\mathrm{FW}\vec r\op{T}_\mathrm{FW}^{-1} 
\end{equation}
and 
\begin{equation}
  \label{eq:r_Pr}
  \op{\vec r}_\mathrm{Pr} = 
  \op{T}_\mathrm{Pr}\vec r\op{T}_\mathrm{Pr}^{-1} 
\end{equation}
with the transformations $\op{T}_\mathrm{FW}$ and $\op{T}_\mathrm{Pr}$
defined in \eqref{eq:UFW} and by
\begin{equation}
  \label{eq:UPr}
  \op T_{\mathrm{Pr}} =
  \begin{pmatrix}
    \mathbb{I}_2 & 0 \\[1ex]
    0 & \I\dfrac{\vec\sigma\cdot\opp}{|\opp|}
  \end{pmatrix}\,.
\end{equation}
Determining an explicit expression for the transformation
\eqref{eq:r_FW} yields the famous Fouldy-Wouthuysen mean position
operator
\cite{Foldy:Wouthuysen:1950:Non-Relativistic_Limit}\footnote{Note that
  due to a typo the explicit expression for the Fouldy-Wouthuysen mean
  position operator given in the publication by Fouldy and Wouthuysen is
  wrong; an error that has propagated through many publications \cite{Ellis:Siopsis:1982:note,Vries:1970:Foldy-Wouthuysen_Transformations}.}
\begin{equation}
  \op{\vec r}_\mathrm{FW} = \vec r + \I\hbar\left(
    \frac{\I\opSigma\times\opp}{2\oppo(\oppo+m_0c)} -
    \frac{\beta(\vec\alpha\cdot\opp)\opp}{2\oppo^2(\oppo+m_0c)} +
    \frac{\beta\vec\alpha}{2\oppo}
  \right)\,,
\end{equation}
which is equivalent to the Newton-Wigner position operator
\cite{Newton:Wigner:1949:Localized_States}, while \eqref{eq:r_Pr} yields
\begin{equation}
  \op{\vec r}_\mathrm{Pr} = \vec r - \hbar
  \begin{pmatrix}
    0 & 0 \\[1ex]
    0 & \dfrac{\vec\sigma\times\opp}{|\opp|^2}
  \end{pmatrix}\,.
\end{equation}
For numerical calculations of expectation values, however, it is more
convenient to utilize the forms \eqref{eq:r_FW} and \eqref{eq:r_Pr}.

Because of symmetry reasons the expectation values of the position
operators $\op{\vec r}_\mathrm{P}$, $\op{\vec r}_\mathrm{FW}$, and
$\op{\vec r}_\mathrm{Pr}$ are zero if these position operators are
applied to the hydrogenic ground states \eqref{eq:hydrogen_real_1} and
\eqref{eq:hydrogen_real_2}.  The second moment, however, does not
vanish.  The variance of a nonrelativistic hydrogenic ground state
scales with $1/Z^2$.  Due to relativistic effects the ground states'
wavepackets shrink even faster.  How fast, depends on the definition of
the position operator, as shown Fig.~\ref{fig:spin}\,b).  In principle,
one can measure the variance of the ground state of a highly charged ion
and compare it to the predictions in Fig.~\ref{fig:spin}\,b) and in this
way determine the correct relativistic position operator and indirectly
also the correct relativistic spin operator.

\section{Electron-spin precession in elliptically polarized light}

In the previous section we argued that the spin of free electrons should
be modeled by the Fouldy-Wouthuysen spin operator.  In the following we
will utilize this operator to study a relativistic spin dynamics, which
originates from a coupling of the electron's spin to the spin of an
electromagnetic wave with elliptical polarization.

The electric and magnetic field components of two elliptically polarized
laser fields propagating into the positive or negative direction of the
$x$~axis are given by
\begin{subequations}
  \label{eq:elliptic_fields}
  \begin{align}
    \label{eq:elliptic_fields_E}
    \vec E_{1,2}(\vec r, t) & = \hat{E} \left(
      \cos\frac{2\pi(x\mp c t)}{\lambda}\vec e_y + 
      \cos\left(\frac{2\pi(x\mp c t)}{\lambda}\pm\eta\right)\vec e_z
    \right) \,,\\
    \label{eq:elliptic_fields_B}
    \vec B_{1,2}(\vec r, t) & = \frac{\hat{E}}{c} \left(
      \mp\cos\left(\frac{2\pi(x\mp c t)}{\lambda}\pm\eta\right)\vec e_y \pm 
      \cos\frac{2\pi(x\mp c t)}{\lambda}\vec e_z
    \right)\,.
  \end{align}
\end{subequations}
Here, the position vector $\vec r =\trans{(x,y,z)}$, the time $t$, and
$\vec e_x$, $\vec e_y$, and $\vec e_z$ denoting unit vectors in the
direction of the coordinate axes are used.  The parameter
$\eta\in(-\pi,\pi]$ determines the degree of the light beams'
ellipticity with $\eta=0$ and $\eta=\pi$ corresponding to linear
polarization and $\eta=\pm\pi/2$ to circular polarization. The two
electromagnetic waves \eqref{eq:elliptic_fields} feature the same
wavelength $\lambda$, the same electric field amplitude $\hat{E}$, and
the same intensity
\begin{equation}
  \label{eq:intensity}
  I = \varepsilon_0c\hat{E}^2 \,,
\end{equation}
but have opposite helicity.  Introducing the wave number
$k=2\pi/\lambda$ and the lasers fields' angular frequency $\omega=kc$,
the Coulomb gauge vector potentials $\vec A_{1,2}(\vec r, t)$ of the
elliptically polarized fields~\eqref{eq:elliptic_fields} are
\begin{equation}
  \vec A_{1,2}(\vec r, t) = -\frac{\hat{E}}{\omega} \left(
    \mp\sin(kx \mp \omega t)\,\vec e_y 
    \mp\sin(kx \mp \omega t\pm\eta)\,\vec e_z
  \right) \,.
\end{equation}
Each of the electromagnetic fields specified by
\eqref{eq:elliptic_fields} carries the photonic spin density
\begin{equation}
  \label{eq:spin_density}
  \varepsilon_0 \vec E_{1,2}\times \vec A_{1,2} =
  \frac{\varepsilon_0 \hat{E}^2\lambda\sin\eta}{2\pi c}\vec{e}_x\,.
\end{equation}

As one can show via the Volkov solution of the Dirac
equation\cite{Wolkow:1935:Uber_Klasse} a single plane wave as given in
\eqref{eq:elliptic_fields} cannot change the spin orientation of an
electron.  Therefore, we consider a standing wave, which is formed by
superimposing the two counterpropagating waves given in
\eqref{eq:elliptic_fields}.  The magnetic vector potential of the
combined laser fields is given by
\begin{equation}
  \label{eq:A}
  \vec A(\vec r, t) = 
  -\frac{2w(t)\hat{E}}{\omega}\cos kx
  \left( 
    \sin \omega t\,\vec e_y + \sin(\omega t-\eta)\,\vec e_z
  \right)\,.
\end{equation}
Here the window function
\begin{equation}
  \label{eq:turn_on_and_off}
  w(t) = 
  \begin{cases}
    \sin^2 \frac{\pi t}{2\Delta T} & \text{if $0\le t\le \Delta T$,} \\
    1 & \text{if $\Delta T\le t\le T-\Delta T$,} \\
    \sin^2 \frac{\pi (T-t)}{2\Delta T} & \text{if $T-\Delta T\le t\le T$,} 
  \end{cases}
\end{equation}
was introduced to allow for a smooth turn-on and turn-off of the laser
field.  The parameters $T$ and $\Delta T$ denote the total interaction
time and the turn-on and turn-off intervals.  For circularly polarized
plane waves ($\eta=\pi/2$), the electric and the magnetic components of
the standing wave are parallel to each other and rotate around the
propagation direction. The maxima of the electric and the magnetic field
components are shifted against eachother by $\lambda/4$.

Solving the time-dependent Dirac equation till time $t=T$ with a common
eigenstate of the free Dirac Hamiltonian, the momentum operator, and the
$z$-component of the Foldy-Wouthuysen spin operator with zero momentum
and positive spin as initial condition shows that the electron's spin
precesses around the propagation axis of the electromagnetic fields
\cite{Bauke:Ahrens:etal:2014:Electron-spin_dynamics,Bauke:Ahrens:etal:2014:Electron-spin_dynamics_2}.
The role of the photonic spin density for the electronic spin precession
becomes evident by considering the weakly relativistic limit of the
Dirac equation \eqref{eq:Dirac_equation}.  In this limit, this equation
reduces via a Foldy-Wouthuysen transformation
\cite{Foldy:Wouthuysen:1950:Non-Relativistic_Limit,Vries:1970:Foldy-Wouthuysen_Transformations,Frohlich:Studer:1993:Gauge_invariance}
to
\begin{multline}
  \label{eq:Dirac_FW_equation}
  \I \hbar\dot \Psi(\vec r, t)  = 
  \Bigg( 
  \frac{( -\I\hbar\vec\nabla - q \vec A(\vec r, t) )^2}{2m_0} -
  \frac{q\hbar}{2m_0}\vec\sigma\cdot \vec B(\vec r, t) 
  +
  q\phi(\vec r, t) 
  -
  \frac{( -\I\hbar\vec\nabla - q \vec A(\vec r, t) )^4}{8m_0^3c^2} -
  \frac{q^2\hbar^2}{8m_0^3c^4}(c^2\vec B(\vec r, t)^2-\vec E(\vec r, t)^2) \\
  -
  \frac{q\hbar}{4m_0^2c^2 }\vec\sigma\cdot (\vec E(\vec r, t) \times (
  -\I\hbar\vec\nabla - q\vec A(\vec r, t) ) ) -
  \frac{q\hbar^2}{8m_0^2c^2}\vec\nabla\cdot\vec{E}(\vec r, t) +
  \frac{q\hbar}{8m_0^3c^2}\acommut{\vec\sigma\cdot\vec B(\vec r, t)}{( -\I\hbar\vec\nabla - q \vec A(\vec r, t) )^2}
  \Bigg)\,\Psi(\vec r, t) 
\end{multline}
\begin{figure}
  \centering
  \includegraphics[scale=0.5]{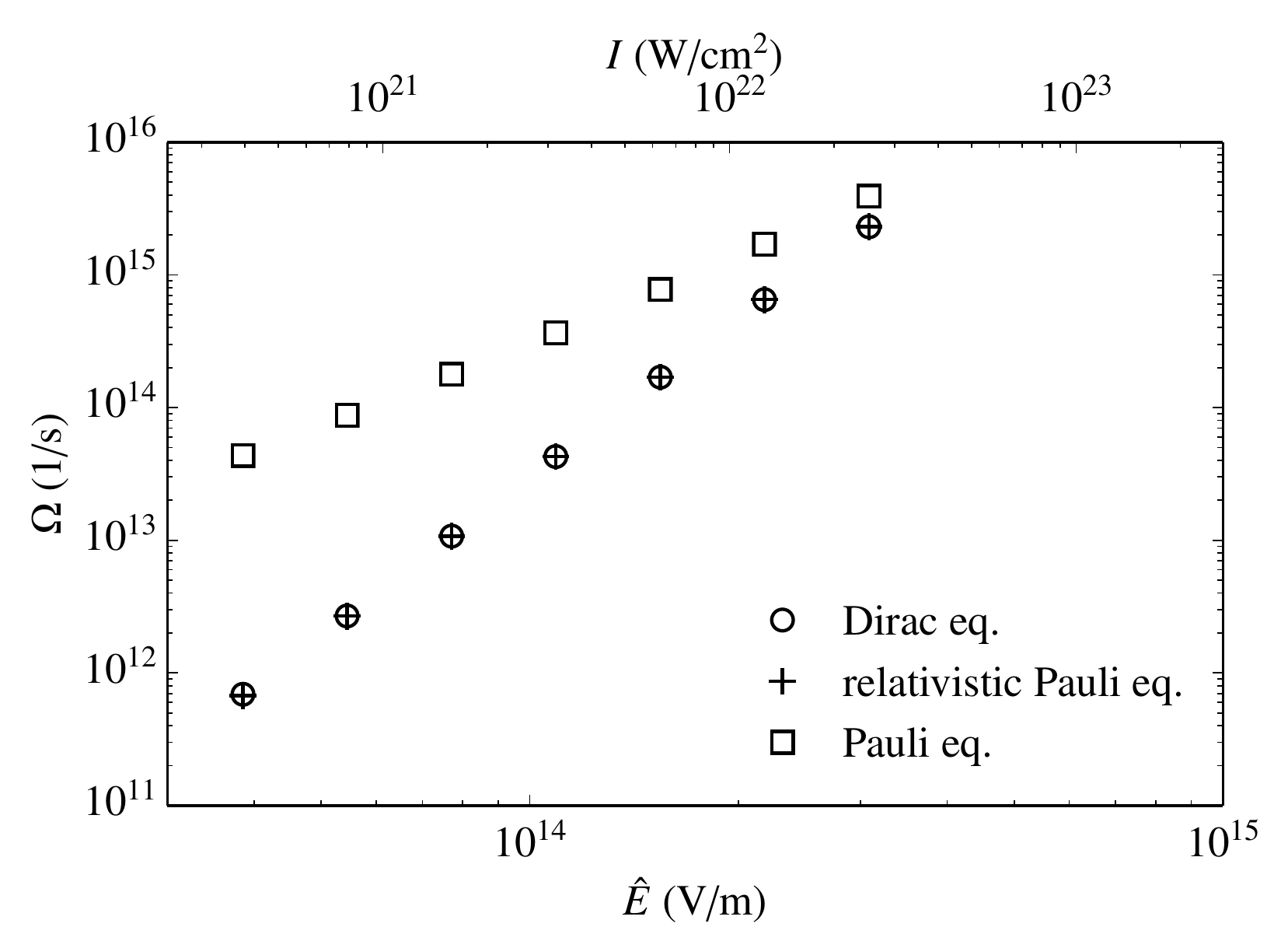}
  \caption{Angular frequency $\Omega$ of the spin precession as a
    function of the laser's electric field strength $\hat{E}$ and its
    intensity $I$ for electromagnetic fields with a wavelength
    $\lambda=0.159\,\mathrm{nm}$.  Depending on the applied theory (the
    Dirac equation \eqref{eq:Dirac_equation}, the relativistic Pauli
    equation \eqref{eq:Pauli_FW_equation}, or the nonrelativistic Pauli
    equation) the spin of an electron in two counterpropagating
    circularly polarized light waves scales with the second (Pauli
    equation) or the fourth power (Dirac equation, relativistic Pauli
    equation) of $\hat{E}$.  Numerical data adopted from {\em New
      J. Phys.}~{\bf 16}(4), 043012 (2014).}
  \label{fig:scaling}
\end{figure}%
for the now two-component wave function $\Psi(\vec r, t)$ with the
vector potential $\vec A(\vec r, t)$ given by \eqref{eq:A} and the
electromagnetic fields $\vec B(\vec r, t)=\vec\nabla\times\vec A(\vec r,
t)$ and $\vec E(\vec r, t)=-\dot{\vec A}(\vec r, t)$.  In leading order
Eq.~\eqref{eq:Dirac_FW_equation} features four terms that may cause spin
dynamics.  The so-called Zeeman term $\sim\vec\sigma\cdot \vec B(\vec r,
t)$ and its the lowest-order relativistic correction given by the
anticommutator expression \mbox{$\sim\acommut{\vec\sigma\cdot\vec B(\vec
    r, t)}{( -\I\hbar\vec\nabla - q \vec A(\vec r, t) )^2}$} mediate the
coupling of the electron's spin to the magnetic field. The term
$\sim\vec\sigma\cdot\vec E(\vec r, t) \times \I\hbar\vec\nabla$ leads to
the so-called spin-orbit interaction, i.\,e., the coupling between the
electron's spin and its orbital angular momentum.  The term
$\sim\vec\sigma\cdot(\vec E(\vec r, t) \times \vec A(\vec r, t))$ may be
interpreted as a coupling of the spin density of the external
electromagnetic wave to the electron's spin.  Considering the
relativistic correction due to the electromagnetic wave's spin density
as the only relativistic correction to the nonrelativistic Pauli
equation the relativistic electron motion in the vector potential
\eqref{eq:A} may be described by the relativistic Pauli equation
\cite{Bauke:Ahrens:etal:2014:Electron-spin_dynamics,Bauke:Ahrens:etal:2014:Electron-spin_dynamics_2}
\begin{equation}
  \label{eq:Pauli_FW_equation}
  \I \hbar\dot \Psi(\vec r, t)  = 
  \Bigg(
  \frac{1}{2m_0}( -\I\hbar\vec\nabla - q \vec A(\vec r, t) )^2 -
  \frac{q\hbar}{2m_0}\vec\sigma\cdot \vec B(\vec r, t)  +
  \frac{q^2\hbar}{4m_0^2c^2 }\vec\sigma\cdot (\vec E(\vec r, t) \times \vec A(\vec r, t) )
  \Bigg)\,\Psi(\vec r, t) \,.
\end{equation}

Numerical calculations indicate that the relativistic Pauli equation
\eqref{eq:Pauli_FW_equation} is sufficient to reproduce the spin
dynamics of the fully relativistic Dirac equation
\eqref{eq:Dirac_equation}.  The nonrelativistic Pauli equation, however,
yields a completely different spin dynamics as shown in
Fig.~\ref{fig:scaling}.  Depending on if the Dirac equation
\eqref{eq:Dirac_equation}, the relativistic Pauli equation
\eqref{eq:Pauli_FW_equation}, or the nonrelativistic Pauli equation is
applied, the spin of an electron in two counterpropagating circularly
polarized light waves scales with the second (Pauli equation) or the
fourth power (Dirac equation, relativistic Pauli equation) of the
electric field's amplitude $\hat{E}$.  Further analytical calculations
based on the Dirac equation and time-dependent perturbation theory
\cite{Bauke:Ahrens:etal:2014:Electron-spin_dynamics_2} show that the
electron's spin precesses with an angular frequency that is proportional
to the photonic spin density $\varrho_\sigma={\varepsilon_0
  \hat{E}^2\lambda}/{(\pi c)}$, the laser field's intensity $I$ given in
\eqref{eq:intensity}, and the fourth power of the wavelength:
\begin{equation}
  \Omega = \varrho_\sigma I \lambda^4 
  \frac{\alpha_\mathrm{el}^2}{2\pi^2m_0^2 c^3}\,.
\end{equation}
The proportionality factor ${\alpha_\mathrm{el}^2}/{(2\pi^2m_0^2 c^3)}$
is independent of the standing wave's electromagnetic field.

\begin{figure}
  \centering
  \includegraphics[scale=0.5]{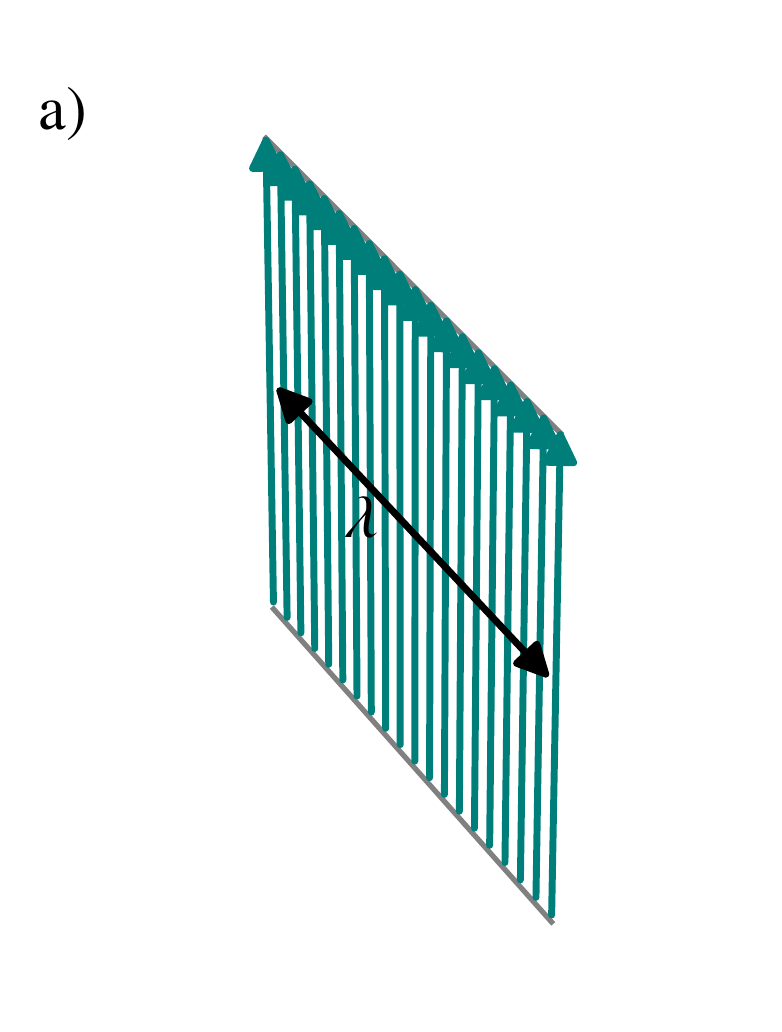}
  \qquad
  \includegraphics[scale=0.5]{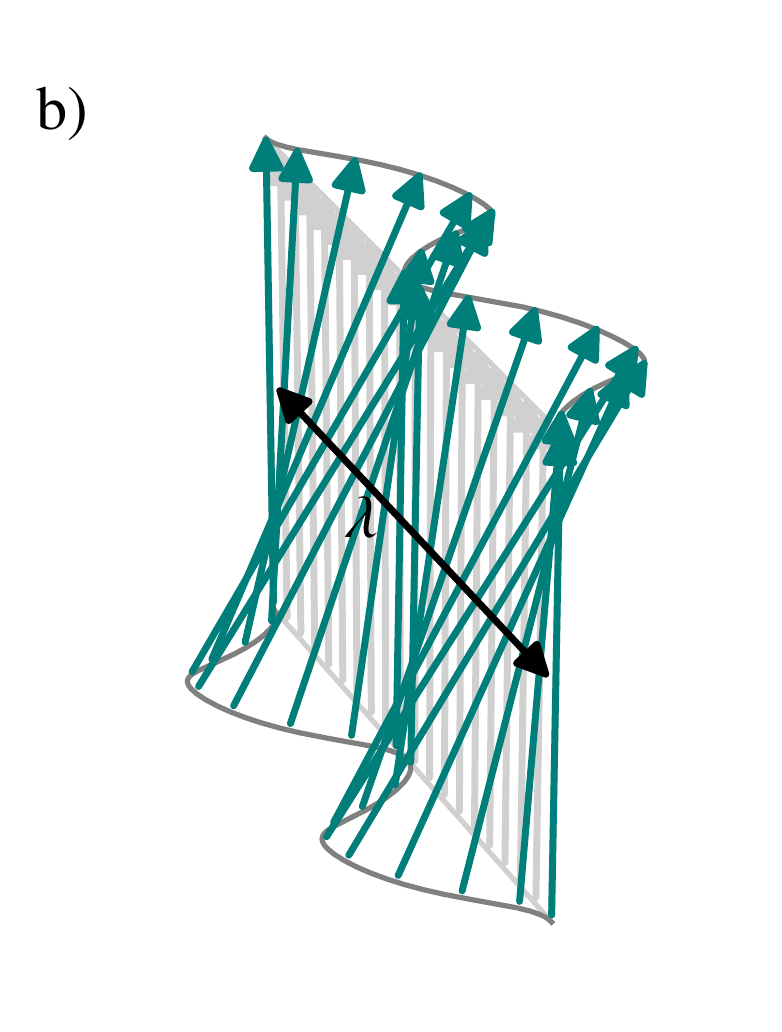}
  \qquad
  \includegraphics[scale=0.5]{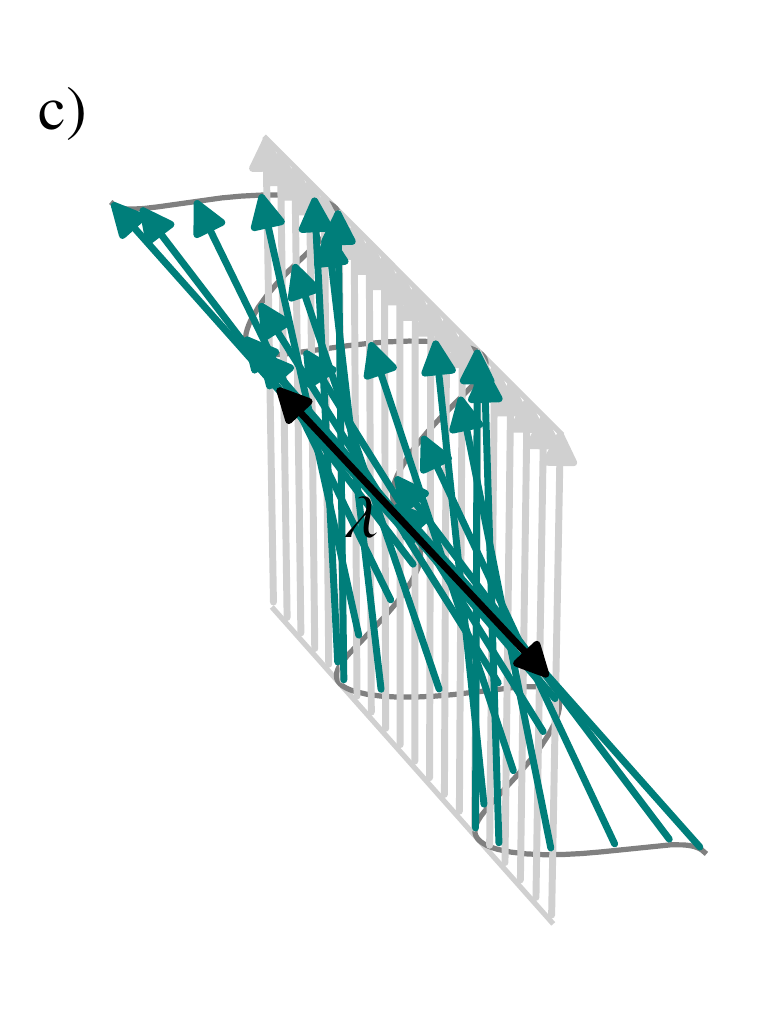}
  \caption{An ensemble of classical spins under the effect of a magnetic
    field and a photonic spin density of a standing wave formed by two
    counterpropagating circularly polarized waves.  Part~a): Initially
    spins are aligned.  Part~b): The rotating magnetic field causes a
    position-dependent rotation of the spin vectors. Part~c): Also the
    photonic spin density causes a spin rotation but into the opposite
    direction.  Averaged over a wavelength both effects cancel
    eachother.}
  \label{fig:classical}
\end{figure}

The role of the relativistic correction due to the photonic spin density
is pivotal, its influence on the scaling of the spin precession
frequency does not become small in the limit of weak fields, which is
usually related to a nonrelativistic limit.  The nonexistence of a
nonrelativistic limit results because the Zeeman term is due to the fast
oscillation of the magnetic field effectively as strong as the
relativistic correction due to the photonic spin density.  On the basis
of a classical argument
\cite{Bauke:Ahrens:etal:2014:Electron-spin_dynamics} one can show that
the effects of the Zeeman term and the correction due to the photonic
spin density on the motion of the electron spin counteract eachother.
The initial electron's quantum state, which is a momentum eigenstate, is
delocalized over several laser wavelengths.  Thus, we mimic the quantum
wavepacket by an ensemble of classical particles with spin angular
momentum.  These particles are placed along the $x$ axis with initially
aligned spin direction, see also Fig.~\ref{fig:classical}\,a).  The
dynamics of a classical electron spin $\vec s$ at fixed position $\vec
r$ in the magnetic field $\vec B(\vec r, t)=\vec B_{1}(\vec r, t)+\vec
B_{2}(\vec r, t)$ is governed by the classical equation of motion
\begin{equation}
  \label{eq:s_B}
  \dot{\vec s}(t) = 
  \frac{q}{m_0} \vec s(t)\times\vec B(\vec r, t) =
  \frac{2q\hat E\cos kx}{m_0} \vec s(t)\times(
    \cos\omega t\,\vec e_y + \sin\omega t\,\vec e_y 
  )\,.
\end{equation}
In a static and homogeneous magnetic field of strength $\hat E/c$, the
electron's spin would precess around an axis parallel to the magnetic
field's direction with the angular frequency $\Omega_\mathrm{L}=q\hat
E/(m_0c)$.  For a rotating magnetic field and parameters such that
$\Omega_\mathrm{L}\ll\omega$, however, the magnetic field rotates so
fast that the spin precesses around the rotation axis of the magnetic
field.  For the magnetic field of a standing wave formed by two
counterpropagating circularly polarized waves given by the vector
potential \eqref{eq:A} a position-dependent angular frequency
$2\Omega_\mathrm{P}\sin^2 kx$ results with $\Omega_\mathrm{P}={(q\hat
  E)^2\lambda}/({2\pi m_0^2c^3})$, see Fig.~\ref{fig:classical}\,b).
Similarly, the relativistic correction due to the photonic spin density
in \eqref{eq:Pauli_FW_equation} to the Pauli equation leads to the
classical equation
\begin{equation}
  \label{eq:s_ExA}
  \dot{\vec s}(t) = -\frac{q^2}{2m_0^2c^2} 
  \vec s(t)\times(\vec E(\vec r, t)\times\vec A(\vec r, t)) =
  -\frac{(q\hat{E})^2\lambda\cos^2 kx}{\pi m_0^2c^3} 
  \vec s(t)\times\vec{e}_x \,.
\end{equation}
This yields also a position-dependent angular frequency, but now
$-2\Omega_\mathrm{P}\cos^2 kx$, see Fig.~\ref{fig:classical}\,c).  As a
consequence, a classical spin under the effect of both spin terms, the
Zeeman term and the relativistic term due to the photonic spin density,
rotates in a short time interval $\Delta t$ around an angle of about
$2\Omega_\mathrm{P}(\sin^2 kx - \cos^2 kx)\Delta t$.  Averaged over a
laser wavelength this rotation angle vanishes and the effects of the two
spin terms chancel each other in our classical model.  Thus, the
classical model explains how the effect of the laser fields' spin
density leads to a breakdown of the quadratic scaling of the
spin-precession angular frequency in $\hat E$ that results if only the
magnetic field is taken into account.  The model, however, is not able
to reproduce the quartic scaling in $\hat E$ that results from the fully
relativistic quantum mechanical Dirac equation.  The quartic scaling
results as a genuine quantum effect from the fast temporal oscillations
combined with the spatial modulation of the electromagnetic fields.  In
fact, further numerical calculations show that the quantum mechanical
wavepacket accumulates in regions of the standing light wave with high
magnetic fields \cite{Bauke:Ahrens:etal:2014:Electron-spin_dynamics_2}.

\ifdefined\Preprint
\else
\makeatletter
\def\thebibliography#1{\section*{REFERENCES\@mkboth
 {REFERENCES}{REFERENCES}}\list
 {[\arabic{enumi}]}
 {\settowidth\labelwidth{[#1]}\leftmargin\labelwidth
 \advance\leftmargin\labelsep \usecounter{enumi}}
 \def\newblock{\hskip .11em plus .33em minus .07em}
  \parskip -0.7ex plus 0.5ex minus 0ex   
\if0\@ptsize\else\small\fi      
 \sloppy\clubpenalty4000\widowpenalty4000
 \sfcode`\.=1000\relax}
\let\endthebibliography=\endlist
\makeatother
\bibliographystyle{spiebib}
\fi
\bibliography{relativistic_spin_dynamics}

\end{document}